\documentclass{article}

\usepackage{arxiv}

\usepackage[utf8]{inputenc} % allow utf-8 input
\usepackage[T1]{fontenc}    % use 8-bit T1 fonts
\usepackage{hyperref}       % hyperlinks
\usepackage{url}            % simple URL typesetting
\usepackage{booktabs}       % professional-quality tables
\usepackage{amsfonts}       % blackboard math symbols
\usepackage{nicefrac}       % compact symbols for 1/2, etc.
\usepackage{microtype}      % microtypography
\usepackage{lipsum}
\usepackage{graphicx}
\graphicspath{ {./images/} }
\usepackage{multirow}
\usepackage{amsmath}
\usepackage{array}
\usepackage{graphicx}
\usepackage{subcaption}
\usepackage{paralist}
\usepackage{tikz}
\usetikzlibrary{shadows}
\usepackage[table]{xcolor}
\usepackage{xspace}
\usepackage{adjustbox}

\newcommand{\tikzcircle}[2][1pt]{%
  \tikz[baseline=(char.base)]%
    \node[draw,circle,inner sep=#1] (char) {#2};%
}

\newcommand{\ds}{dataset\xspace}
\newcommand{\dss}{datasets\xspace}

\title{Rethinking Accuracy: \\ A Weighted Error-Based Metric for Data Quality}

\newbox{\orcid}\sbox{\orcid}{\includegraphics[scale=0.06]{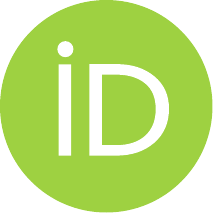}}
\author{
  \href{https://orcid.org/0000-0002-5960-5886}{\usebox{\orcid}\hspace{1mm}Valerie Restat} \\
  FernUniversität in Hagen \\
  Hagen, Germany\\
  \texttt{valerie.restat@fernuni-hagen.de} \\
  \And
  \href{https://orcid.org/0000-0003-2771-142X}
   {\usebox{\orcid}\hspace{1mm}Uta St{\"o}rl} \\
  FernUniversität in Hagen  \\
  Hagen, Germany\\
  \texttt{uta.stoerl@fernuni-hagen.de} \\
}

\begin{document}
\maketitle
\begin{abstract}
Real data often contains errors, which is why data engineers spend a lot of time creating data cleaning pipelines to ensure the best possible data quality. However, it is often difficult to compare the results of different pipelines and decide which pipeline leads to the best results. There are many different metrics that are designed for different use cases, but they often take only a portion of the data into account. There is a lack of universally applicable metrics for measuring data quality that can be used in many different scenarios. That is why in this paper we are presenting TOMME -- an initial approach to a universally applicable weighted error-based metric for data quality. This allows the data quality of a dataset to be assessed based on a single score. While a detailed data quality evaluation remains important, the use of a single score enables rapid assessment and automated processing, for example, for optimization algorithms. By using different weights, the score can also be precisely adjusted to the specific use case. That is why we named it TOMME, which stands for ``The One Metric Measuring Errors''. As the name suggests, it measures errors in the data. It can thus be considered a generalized, weighted form of accuracy.
\end{abstract}

% keywords can be removed

\section{Introduction}
Imagine you are a data engineer and you create data cleaning pipelines to improve the quality of your \dss. You implement different pipelines and want to compare the different results with each other. While a detailed report listing individual errors is important to you, you also need a metric that combines these values so that you can see at a glance which pipeline delivers better results for you. Moreover, depending on the use case, different aspects may be important. In one application, it is particularly important that there are no missing values. In another application, missing values are less significant, but there must be no syntax violations. The properties of your \dss also vary in importance. This is where TOMME comes to the rescue: It offers a single metric to combine different error measures. 

The basis is provided by a detailed error report. We use CheDDaR~\cite{CheDDaR} for this purpose. CheDDaR is a framework for detecting data quality issues. It is flexible and expandable and includes various verification methods. As an alternative to CheDDaR, other error detection tools that provide a correspondingly detailed report can also be used. TOMME builds on this report and allows many different measures to be combined. In addition, weights can be specified depending on what is important for the specific use case. This enables comparisons between different data cleaning pipelines. While metrics such as accuracy and F1 score have become established in the context of machine learning evaluation, to the best of our knowledge there are no such established metrics for assessing data quality. This is also pointed out by Sambasivan et al.~\cite{Sambasivan2021}, who state that the current focus is on metrics for measuring ``goodness of the fit of the model to the data'' -- as measured by F1, accuracy, etc. The authors emphasize that metrics for measuring ``goodness-of-data'' are needed instead -- as we are aiming for with TOMME. The lack of measurability of data quality has already been described by Sebastian-Coleman~\cite{Sebastian2012} in 2013 as a major challenge, and Ehrlinger et al.~\cite{Ehrlinger2025} argue that this is still true even in the age of artificial intelligence. With TOMME, we want to offer an initial approach to solving this problem.

TOMME allows the quality of \dss to be measured. By comparing the data quality of the \dss before and after a run of a data cleaning pipeline, it is also possible to determine the quality of pipelines and optimize them. Moreover, TOMME is an important contribution to \textit{automatic} optimization of data quality. In most cases, the search space of all possible data cleaning pipelines is too large to be examined manually by a domain expert or using a brute-force approach. Instead, a cost function is needed for optimization (as we described in FONDUE~\cite{FONDUE}, for example). The metric determined with TOMME can also be used for this purpose.

The contribution of this paper is as follows:
\begin{itemize}
    \item We present a formula for measuring the ``goodness-of-data''.
    \item The metric produces a single score.
    \item We are introducing a system that simplifies the application of the metric via a web UI. 
\end{itemize}

The remainder of the paper is structured as follows: In Section~\ref{sec:sota} we present related work. Section~\ref{sec:ch+t} then explains the formula of TOMME. To illustrate how it works, three experiments are conducted in Section~\ref{sec:exp}. Section~\ref{sec:usage} describes the system we built around TOMME to simplify its use. Finally, the paper is summarized in Section~\ref{sec:conc}, and an outlook on future work is provided.

\section{Related Work}
\label{sec:sota}
Data quality is a well-researched field and various metrics already exist~\cite{Wang1995, Blake2011, Heinrich2018, Azeroual2018, Elouataoui2022}. However, these cover only part of data quality or aim at specific applications. Since data quality encompasses so many different aspects, it is usually divided into different dimensions, with \textit{accuracy} as one of the most important data quality constructs~\cite{Haegemans2016}. However, there are different definitions and no standardized measures~\cite{Ehrlinger2025}. Haegemans et al.~\cite{Haegemans2016} analyze that the most commonly used notion of accuracy is the \textit{magnitude of an error}. This applies both to single data items and multiple data items. However, they also show that the measurement is still conducted differently, often according to the \textit{occurrence of an error}, even though this contradicts the original notion~\cite{Haegemans2016}. The most widely used metric for accuracy, which can be applied at different levels of granularity, is as follows~\cite{Ehrlinger2025}: 
\newline
% \centerline{
accuracy = $\frac{\text{number of correct data items}}{\text{number of all data items}}$
% }
%accuracy = (number of correct data items) / (number of all data)
% \begin{equation}
%     \text{accuracy} = \frac{\text{number of correct data items}}{\text{number of all data items}}
% \end{equation}

However, the formula we present with TOMME can be configured in more detail than this approach. The weights for error types and properties allow it to be customized to your specific use case. It is thus based on the \textit{magnitude of an error}, as defined in the main definition. Because all error types are taken into account and the formula is not limited to the data item level, overlapping errors and different data formats in addition to tabular data (e.g., semi-structured data) can be considered. That is why we name it TOMME, which stands for ``The One Metric Measuring Errors'' and it can be seen as a a generalized, weighted form of accuracy.

Various tools are available for detecting errors and validating data. Commonly used open source tools include pydantic\footnote{\url{https://github.com/pydantic/pydantic}}, pandera\footnote{\url{https://github.com/unionai-oss/pandera}}, and Great-expectations\footnote{\url{https://github.com/great-expectations/great_expectations}}. In research, there are systems such as Raha~\cite{Mahdavi2019} and HoloDetect~\cite{Heidari2019}. These use machine learning to reduce the manual effort involved in error detection. In contrast, other approaches such as the one by Shresta et al.~\cite{Shrestha2023} and MetricDoc~\cite{Bors2018} focus specifically on manual evaluation by domain experts. With CheDDaR, we provide a domain-independent framework. Different verification methods allow for a wide range of applications, depending on how much domain knowledge is available.

Schelter et al.~\cite{Schelter2019} have demonstrated Deequ, a system to evaluate data quality. They show how users can define unit tests for data and how Deequ can be used to validate growing \dss. Abdelaal et al.~\cite{Abdelaal2025} have demonstrated DataLens, an interactive dashboard for managing the data quality of tabular data in the context of machine learning. It includes methods for data profiling, error detection and repair, as well as various machine learning methods. 

Current research also focuses on the use of LLMs for error detection and cleaning~\cite{Ni2024, Biester2024}. However, Bodensohn et al. have shown that these models do not work well with enterprise data~\cite{Bodensohn2024}. In addition, the use of LLMs often involves very high resource consumption~\cite{Patterson2021, Lin2023}.

In contrast to the approaches described, that create detailed error reports, our solution uses a metric to calculate a single score to evaluate data quality. While detailed error reports remain important, such reports do not provide a metric for assessing ``goodness-of-data''. This is what we aim to achieve with TOMME.

\section{Formula}
\label{sec:ch+t}
As mentioned, TOMME stands for ``The One Metric Measuring Errors'' and it can be seen as a a generalized, weighted form of accuracy. As the name suggests, it measures errors in the data. The basis for this is a detailed error report. From the individual error measures provided by such a report, the metric of TOMME can then be used to calculate a \textit{data quality score}. We call this score $DQ_{\mathrm{TOMME}}$. The following describes the derivation of the formula used to calculate this data quality score. 

A \ds consists of several \texttt{properties} $p_i \in P, i \in \{1, ..., |P|\}$. In tabular data, these are columns. However, we use the standardized term property~\cite{Koupil2022} to refer not only to relational data. There may be various \texttt{errors} $e_i \in E, i \in \{1, ..., |E|\}$ in a \ds. Errors are classified according to an \texttt{error type} $et$, e.g., \textit{missing value} or \textit{outlier}.

First, we look at the base metric called $BaseDQ^{\mathrm{TOMME}}$. This takes into account all errors found in the data. Each \texttt{error type} and each \texttt{property} can be weighted individually, depending on the use case. The formula is as follows:
\begin{equation}
    BaseDQ^{\mathrm{TOMME}} = 1- \frac{\sum_{e=1}^{|E|} w^{et}_{e} \cdot w^{p}_{e} \cdot q_e}{\sum_{e=1}^{|E|}w^{et}_{e} \cdot w^{p}_{e}}
\end{equation}

Where:
\begin{equation}
\begin{split}
    w^{et} : \text{weight of \texttt{error type} } et \\
    w^{p} : \text{weight of \texttt{property} } p \\
    q_e \in [0, 1] : \text{value of \texttt{error} } e \\
    |E| : \text{number of \texttt{errors} } \\
    %|P| : \text{number of \texttt{properties}} \\
\end{split}
\end{equation}

$q_e$ is the value of the respective error $e$, where $0$ means that this error occurs 0\% of the time, and $1$ means that 100\% of the values represent an error. For example, if a property \texttt{name} has 10\% missing values, then $q_e = 0.1$. $|E|$ represents the total number of errors in the data. For example, if a dataset has 10\% missing values in the \texttt{name} property, 5\% missing values in the \texttt{age} property as well as 5\% interval violations in \texttt{age}, then $|E| = 3$.  Intuitively, one might expect that the ratio of errors to the total amount of data could be used here. However, this is not possible. For one thing, errors can overlap, and for another, we do not know what kinds of errors might occur in the data.

The weights are defined as follows: By default, the weights for \texttt{error type} and \texttt{property} are $1$:
\begin{equation}
\begin{split}
    w^{et} = 1 \\
    w^p = 1 \\
\end{split}
\end{equation}

However, depending on the use case, it is also possible to select custom weights. These are defined as follows:
\begin{equation}
\begin{split}
    w^{et} = cw^{et} * \frac{|cw^{et}|}{\sum cw^{et}} \\
    w^p = cw^p * \frac{|cw^p|}{\sum cw^p}
\end{split}
\end{equation}

Where:
\begin{equation}
\begin{split}
    cw^{et} \in \mathbb{N}_0: \text{custom weight of \texttt{error type} }et \\
    cw^p \in \mathbb{N}_0: \text{custom weight of \texttt{property} }p
\end{split}
\end{equation}

Thus, weights are always scaled in relation to 1. Due to normalization using the denominator, the metric always assumes a value between 0 and 1 ($dq \in [0,1]$), where 1 is the best possible data quality and 0 is the lowest.

It should be noted that if a weight is set to 0, it is assumed that this error type or property is not relevant for calculating data quality. For this reason, the corresponding values are not included in the number of errors $|E|$.

However, the base metric is not yet sufficient. Intuitively, the more errors there are in the data, the worse the data quality becomes. Since we do not know the maximum possible number of errors, the number of errors is taken into account via a penalty term.

\newpage
There are various options for the penalty term:
\begin{itemize}
    \item \textbf{Linear}: $BaseDQ^{\mathrm{TOMME}} * (1 - \beta \cdot |E|)$
    \item \textbf{Inverse}: $\frac{BaseDQ^{\mathrm{TOMME}}}{1  + |E|}$
    \item \textbf{Logarithmic}: $BaseDQ^{\mathrm{TOMME}} \cdot (1 - \beta \cdot \log(|E| + 1))$
\end{itemize}

The parameter $\beta$ allows the penalty to be flexibly scaled:
\begin{itemize}
    \item Lenient penalty: $\beta \approx 0.05$
    \item Moderate penalty: $\beta \approx [0.1,0.15]$
    \item Strong penalty: $\beta \approx 0.2$
\end{itemize}

\begin{figure}[ht]
    \centering
    \includegraphics[width=0.7\linewidth]{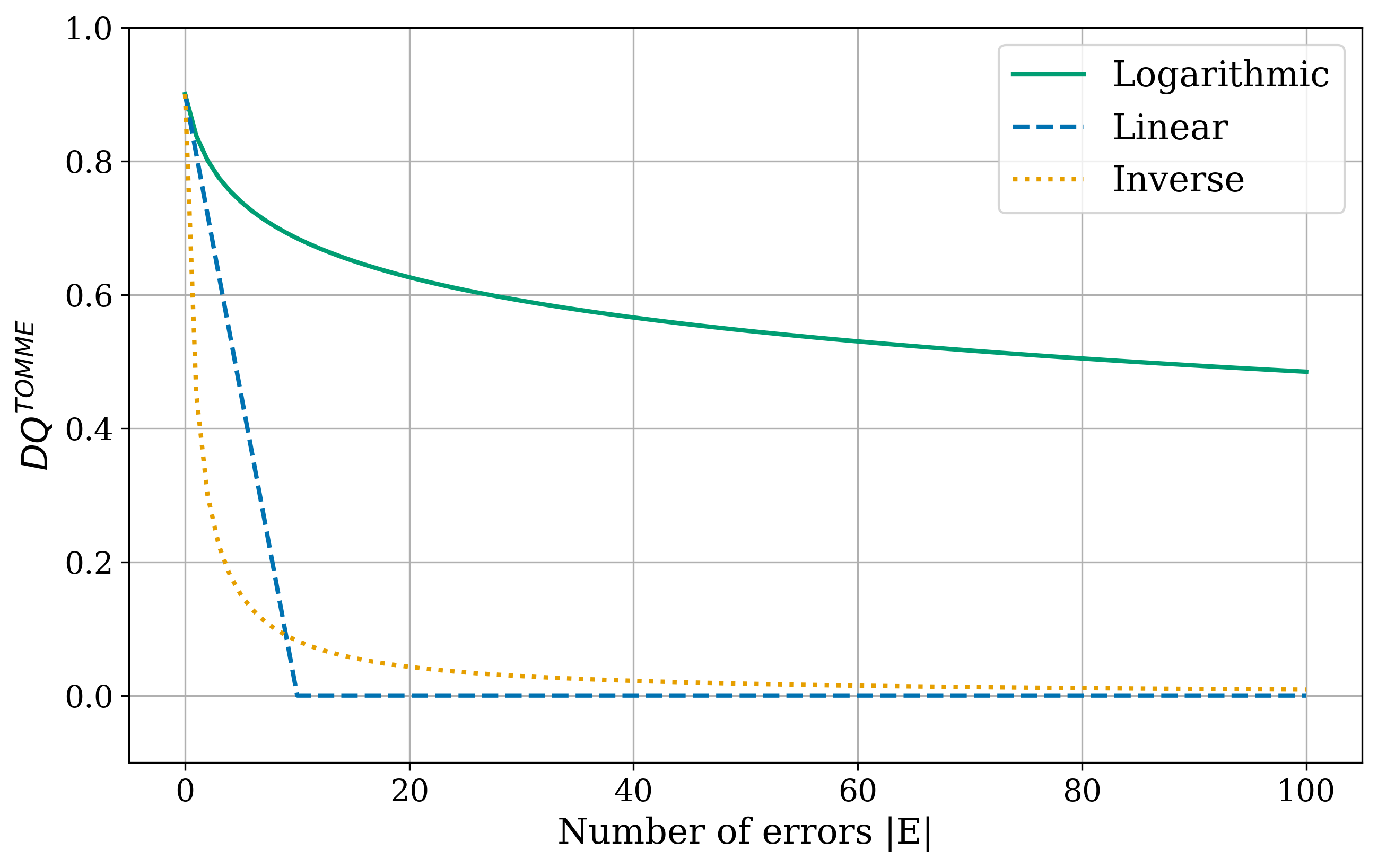}
    \caption{Comparison of the different penalty terms \textit{logarithmic}, \textit{linear}, and \textit{inverse} with $BaseDQ^{\mathrm{TOMME}} = 0.9$ and $\beta = 0.1$}
    %\Description{Line Chart showing comparison of the different penalty terms}
    \label{fig:penatly-comparison}
\end{figure}

Figure~\ref{fig:penatly-comparison} shows the three variants for the penalty term with $BaseDQ^{\mathrm{TOMME}} = 0.9$ and $\beta = 0.1$. It can be seen that linear penalty, even though it is the simplest, very quickly leads to very poor data quality. This is not realistic for large \dss. The values would actually become negative but have been clipped to~0. The inverse variant is slightly milder but even a few errors are penalized very significantly. In contrast, the logarithmic variant offers a smoother decline. Thus, minor errors are not penalized excessively. Even with many errors, the score does not fall to 0, but asymptotically approaches a lower limit. This is more in line with reality.  For this reason, the logarithmic variant was chosen.

\begin{figure}[ht]
    \centering
    \includegraphics[width=0.7\linewidth]{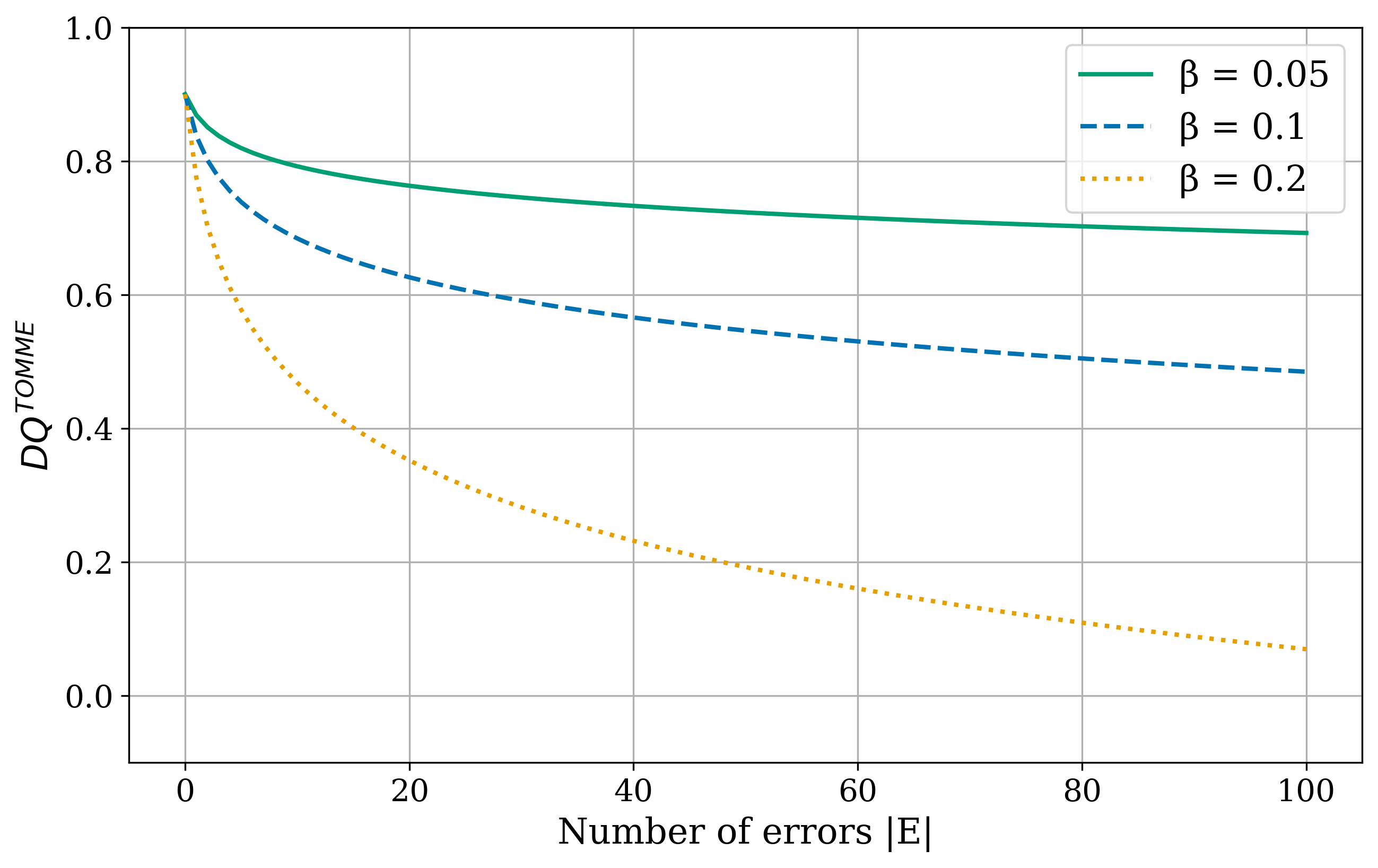}
    \caption{Comparison of the logarithmic penalty terms with different values for $\beta$ with $BaseDQ^{\mathrm{TOMME}} = 0.9$}
    %\Description{Line Chart showing comparison of the different logarithmic penalty terms}
    \label{fig:beta-comparison}
\end{figure}

The effects of the different values of $\beta$ are shown in Figure~\ref{fig:beta-comparison}.

The overall data quality score can accordingly be calculated as follows:

\begin{align*}
DQ^{\mathrm{TOMME}} &= BaseDQ^{\mathrm{TOMME}} \cdot Penalty \\
   &= \left(1 - \frac{\sum_{e=1}^{|E|} w^{et}_{e} \cdot w^{p}_{e} \cdot q_e}{\sum_{e=1}^{|E|} w^{et}_{e} \cdot w^{p}_{e}}\right) \cdot \left(1 - \beta \cdot \log(|E| + 1)\right)
\end{align*}

Additionally, the size of the \ds can be taken into account:
\[DQ^{\mathrm{TOMME}}_{scaled} = \frac{|P_c| + |P_e| \cdot DQ}{|P|}\]

Where $|P_c|$ are the number of correct properties, $|P_e|$ the number of erroneous properties, and $|P|$ the total number of properties. This is suitable for use cases with data that contain a large number of properties. If there are a lot of properties that do not contain errors, this can be taken into account in the metric. As a result, \dss that have many (error-free) properties receive a better score than those that have fewer properties, even if the total number of errors in the data is the same.

In summary, TOMME can be used to calculate a data quality score to compare different \dss. The metric can be customized to suit individual use cases by applying different weights and other parameters. Consequently, TOMME does not claim to be a metric that can be used to compare independent use cases with each other. Rather, its many customization options help you identify the best possible data cleaning pipeline for your specific use case. Thus, it can be used universally for different applications.

\paragraph{Limitations}
CheDDaR is based on the error classification we presented in~\cite{GouDa}. Currently, only errors from level~1 and~2, i.e., those related to individual properties, are taken into account. In the future, both CheDDaR and TOMME will be expanded to include error types from additional error levels, e.g.\ noise. We also want to create an option where the measures from the ISO/IEC 25024:2015~\cite{iso25024} can be used as input (which are currently only partially included). Nevertheless, common errors~\cite{GouDa} have already been taken into account, which is why TOMME can be used in a wide range of applications.

\section{Experiments}
\label{sec:exp}
To demonstrate that the metric described can be used to measure data quality, we conducted experiments on a synthetic \textit{cheese \ds}. Excerpts from this \ds are shown in Table~\ref{tab:cheese}. The \ds contains information about cheese purchases. As shown in Table~\ref{tab:cheese}, there are missing values in the property \texttt{type} and outliers in \texttt{price}. The property \texttt{amount} contains missing values as well.

\begin{table}[ht]
\centering
\caption{Sample excerpt from \textit{cheese \ds}; errors are colored in orange}
\label{tab:cheese}
\begin{adjustbox}{max width=\columnwidth}
\begin{tabular}{l|l|l|r|r}
\toprule
\textbf{idx} & \textbf{name} & \textbf{type} & \textbf{price}& \textbf{amount} \\
\midrule
0 & Crottin de Chavignol & Soft & 30.61  & \cellcolor{orange!25} \\
\midrule
1 & Appenzeller & Hard & 10.32 & 1\\
\midrule
2 & Crottin de Chavignol & \cellcolor{orange!25} & 18.00  & 4\\
\midrule
3 & Boursin & Soft & 16.82  & 2\\
\midrule
4 & Torta del Casar & Soft & \cellcolor{orange!25} 266.81  & 4\\
\midrule
5 & Bleu du Vercors-Sassenage & Blue & 38.58 & 9 \\
\midrule
6 & Garrotxa & Semi-Hard & \cellcolor{orange!25} 331.20  & 3\\
\midrule
7 & Jarlsberg & Semi-Hard & 13.16  & 2\\
\midrule
8 & Sbrinz & Hard & 14.45 & 3 \\
\midrule
9 & Tetilla & \cellcolor{orange!25} & 26.77 & \cellcolor{orange!25} \\
\bottomrule
\end{tabular}
\end{adjustbox}
\end{table}

For the experiments, various versions of this \ds were created, into which errors were inserted in a controlled manner. The data generator GouDa\footnote{\url{https://gitlab.com/cheese-board/gouda}}~\cite{GouDa} was used for this purpose. GouDa's advantage is that it can be used to generate data with and without errors. The error rate can be adjusted flexibly, making it well suited to our use case. 

For the experiments, a value of $\beta=0.1$ was chosen, because we wanted to opt for a moderate penalty. The following subsections provide a more detailed description of the experiments conducted and the datasets generated for this purpose.

\newpage
\subsection{\textit{Experiment A}: Steadily increasing error rates in four properties}
For this experiment, the following errors were introduced into the data:
\begin{itemize}
    \item \textit{Missing values} in \texttt{type}
    \item \textit{Outliers} in \texttt{price}
    \item \textit{Missing values} in \texttt{amount}
\end{itemize}

The error rate remains constant across all properties and increases steadily by 10\%, starting at 0\% (error-free) and rising to 100\% (all errors) -- see Table~\ref{tab:datasets-expA}. It is expected that for $DS_0$, i.e., the error-free \ds, the data quality score will be 1. As the error rate increases, the score will decrease until it finally reaches 0 for $DS_{100}$, the \ds with 100\% errors on the erroneous properties.

\begin{table}[ht]
\centering
\caption{Datasets for \textit{Experiment A} -- the error rate for missing values (MV) and outliers is specified per property for each \ds}
\label{tab:datasets-expA}
\begin{tabular}{ll|ccc}
\toprule
 \multicolumn{2}{c|}{\multirow{2}*{\textbf{\ds}}}
& \multicolumn{3}{c}{\textbf{Error rates}} \\
\cmidrule(lr){3-5}
& 
& MV \texttt{type} 
& Outliers \texttt{price} 
& MV \texttt{amount} \\
\midrule
\multirow{11}*{$|E|=3$}
& $DS_{0}$   & 0.0 & 0.0 & 0.0 \\
& $DS_{10}$  & 0.1 & 0.1 & 0.1 \\
& $DS_{20}$  & 0.2 & 0.2 & 0.2 \\
& $DS_{30}$  & 0.3 & 0.3 & 0.3 \\
& $DS_{40}$  & 0.4 & 0.4 & 0.4 \\
& $DS_{50}$  & 0.5 & 0.5 & 0.5 \\
& $DS_{60}$  & 0.6 & 0.6 & 0.6 \\
& $DS_{70}$  & 0.7 & 0.7 & 0.7 \\
& $DS_{80}$  & 0.8 & 0.8 & 0.8 \\
& $DS_{90}$  & 0.9 & 0.9 & 0.9 \\
& $DS_{100}$ & 1.0 & 1.0 & 1.0 \\
\bottomrule
\end{tabular}
\end{table}

\begin{figure}[ht]
    \centering
    \includegraphics[width=\linewidth]{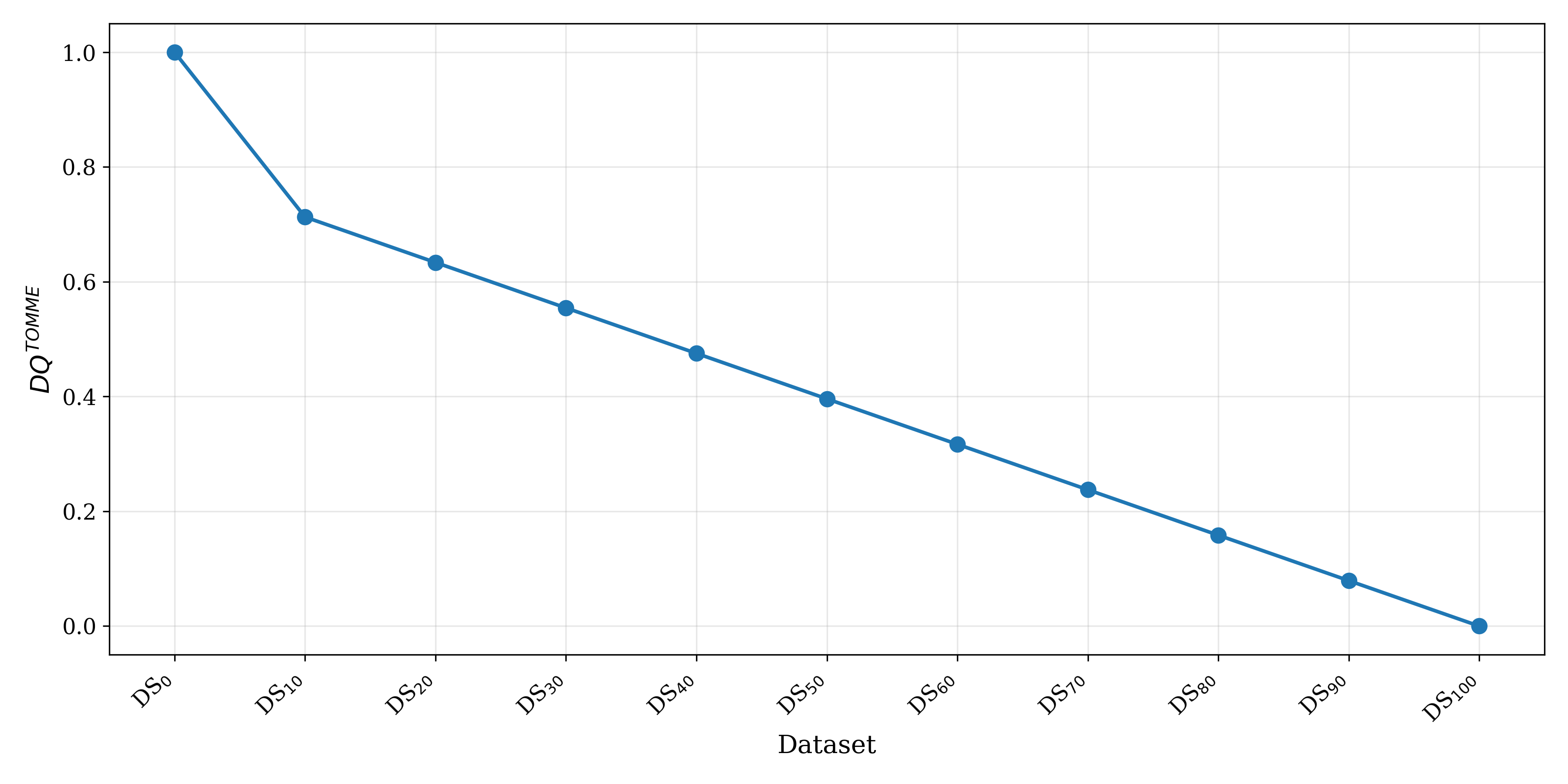}
    \caption{Data Quality Scores for \textit{Experiment A}}
    %\Description{Data Quality Scores for Experiment A}
    \label{fig:dqplot-expA}
\end{figure}

Figure~\ref{fig:dqplot-expA} shows the data quality score across all the \dss described. As expected, the score for $DS_0$ is $1$. Since there are no errors in the data here yet, the penalty term is also $0$. This changes as soon as the data contains errors (starting with $DS_{10}$). That explains the kink at $DS_{10}$. From this point on, the number of errors -- and thus the penalty term -- does not change, and we see a monotonically decreasing incline up to $DS_{100}$, where the score yields $0$ as expected. 

\subsection{\textit{Experiment B}: Steadily increasing error rates across a growing number of properties}
The next experiment is similar to \textit{Experiment A}. Here, too, the error rate increases steadily by 10\%. However, there are now three variations:
\begin{itemize}
    \item $|E|=1$: Only the property \texttt{type} contains errors
    \item $|E|=2$: The properties \texttt{type} and \texttt{price} contain errors
    \item $|E|=3$: The properties \texttt{type}, \texttt{price}, and \texttt{amount} contain errors -- just as in \textit{Experiment A}
\end{itemize}

This experiment is designed to examine the effects of the number of errors $|E|$. It is expected that, for all three test scenarios, the line will follow the same pattern as in \textit{Experiment A}: starting at $1$, then a slight dip, and then a linear trend until the score reaches $0$ at $DS_{100}$. However, the higher the number of errors $|E|$, the lower the score should be.

\begin{figure}[ht]
    \centering
    \includegraphics[width=\linewidth]{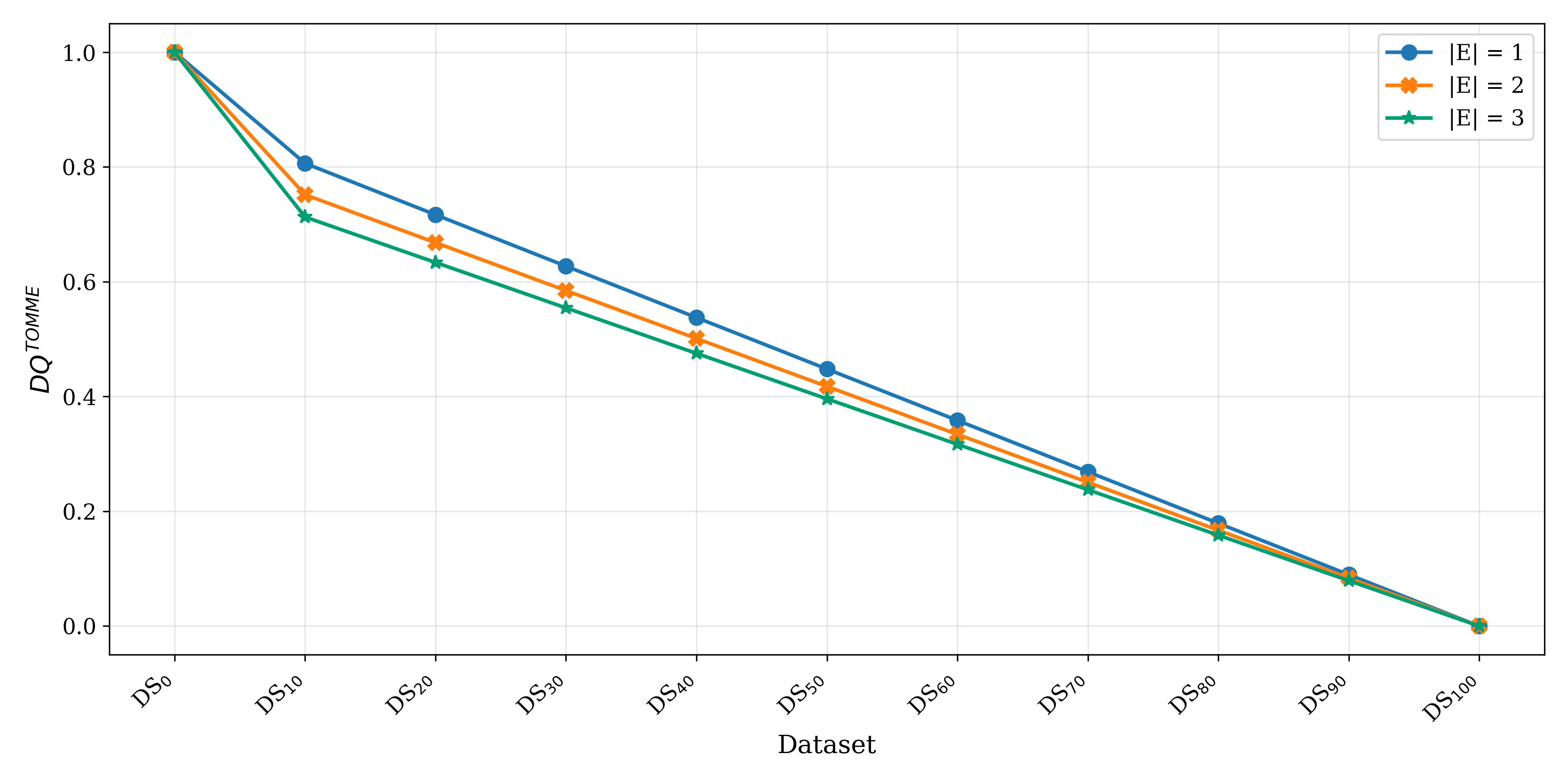}
    \caption{Data Quality Scores for \textit{Experiment B}}
    %\Description{Data Quality Scores for Experiment B}
    \label{fig:dqplot-expB}
\end{figure}

Figure~\ref{fig:dqplot-expB} shows that the results are as expected. Since the logarithmic penalty term with a beta of 0.1 was chosen, the lines are not evenly spaced. As described, this follows the concept of moderate penalization for the number of errors $|E|$. Otherwise, the score would quickly drop to $0$.

\subsection{\textit{Experiment C}: Opposing error rates}
\textit{Experiment C }is designed to evaluate the effects of the weights. To this end, 10 datasets were created, each containing two type of errors:
\begin{itemize}
    \item \textit{Missing Values} in \texttt{type}
    \item \textit{Outliers} in \texttt{price}
\end{itemize}

However, this time the error rates do not rise steadily, but follow the opposite pattern. If 10\% of the values in \texttt{type} are missing, then 90\% of the values in \texttt{price} are outliers, and vice versa (see Table~\ref{tab:datasets-expC}).

\begin{table}[ht]
\centering
\caption{Datasets for \textit{Experiment C} -- the error rate for missing values (MV) and outliers is specified per property for each \ds}
\label{tab:datasets-expC}
\begin{tabular}{ll|cc}
\toprule
 \multicolumn{2}{c|}{\multirow{2}*{\textbf{\ds}}}
& \multicolumn{2}{c}{\textbf{Error rates}} \\
\cmidrule(lr){3-4}
& 
& MV \texttt{type} 
& Outliers \texttt{price} \\
\midrule
\multirow{9}*{$|E|=2$}
& $DS_{10-90}$  & 0.1 & 0.9 \\
& $DS_{20-80}$  & 0.2 & 0.8 \\
& $DS_{30-70}$  & 0.3 & 0.7 \\
& $DS_{40-60}$  & 0.4 & 0.6 \\
& $DS_{50-50}$  & 0.5 & 0.5 \\
& $DS_{60-40}$  & 0.6 & 0.4 \\
& $DS_{70-30}$  & 0.7 & 0.3 \\
& $DS_{80-20}$  & 0.8 & 0.2 \\
& $DS_{90-10}$  & 0.9 & 0.1 \\
\bottomrule
\end{tabular}
\end{table}

The metric is now calculated in three different ways using these \dss:
\begin{itemize}
    \item $w_{mv}=1; w_o=1$: The weights for \textit{missing values (mv)} and \textit{outliers (o)} are equal.
    \item $w_{mv}=2; w_o=1$: \textit{Missing values (mv)} are weighted twice as much as \textit{outliers (o)}.
    \item $w_{mv}=1; w_o=2$: \textit{Outliers (o)} are weighted twice as much as \textit{missing values (mv)}.
\end{itemize}
It is expected that the score will remain constant when the weights are the same, since the opposing error rates balance each other out. In the other two cases, the score is expected to behave exactly oppositely. Thus, for example, the score for \ds $DS_{10-90}$  with weights $w_{mv}=1; w_o=1$ should be the same as for \ds $DS_{90-10}$ with weights $w_{mv}=1; w_o=2$.

\begin{figure}[ht]
    \centering
    \includegraphics[width=\linewidth]{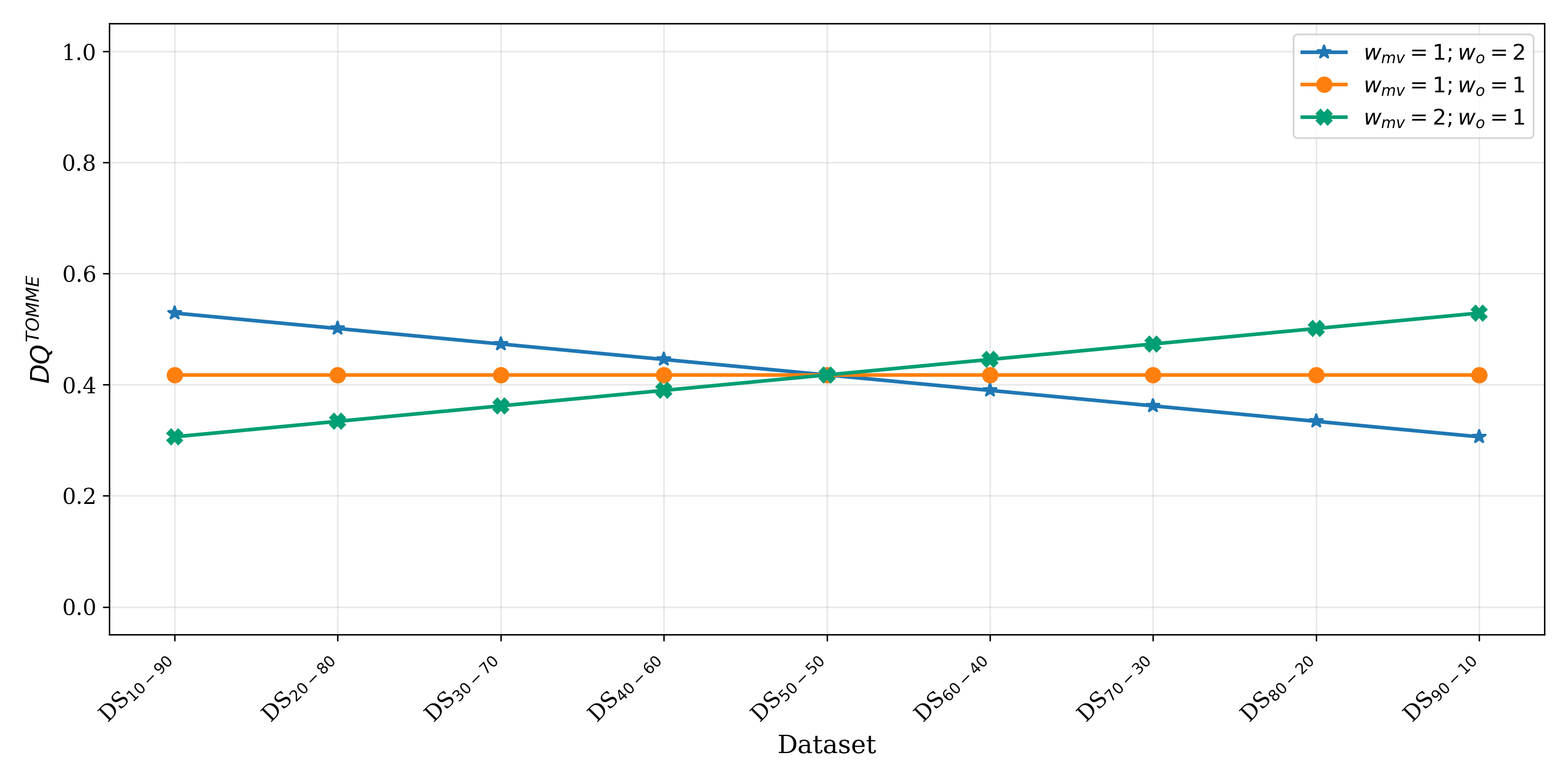}
    \caption{Data Quality Scores for \textit{Experiment C}}
    %\Description{Data Quality Scores for Experiment C}
    \label{fig:dqplot-expC}
\end{figure}

Figure~\ref{fig:dqplot-expC} shows the data quality score of this experiment. The results are as expected: For the equal weights ($w_{mv}=1; w_o=1$), the score remains constant across all \dss. It is slightly lower than $0.5$ because $|E|=2$ is included in the penalty term. The score in the test cases where a particular error type was weighted twice show opposite trends. For $DS_{50-50}$, all scores are the same, since the error rates are the same and therefore the weights have no effect.

Alternatively, the properties can be weighted accordingly, or the weights can be scaled differently (e.g., $w_{mv}=10; w_o=5$). However, since this leads to exactly the same results, it is not shown here.

\section{System}
\label{sec:usage}
The TOMME metric is versatile and can be used in a wide range of applications. The only basis for this is information on errors in the data. In theory, all you need is a pen and paper to calculate the score. However, in the following, we would like to introduce a web UI that makes TOMME even easier to use. We have made the code available at Gitlab\footnote{\url{https://gitlab.com/cheese-board/tomme}} so that the application can be tested. All error profiles from previous experiments are included, so that all measurements can be reproduced.

Figure~\ref{fig:overview} shows an overview of the process. In the following, we will go through each individual step:

\begin{figure}[ht]
    \centering
    \includegraphics[width=0.85\linewidth]{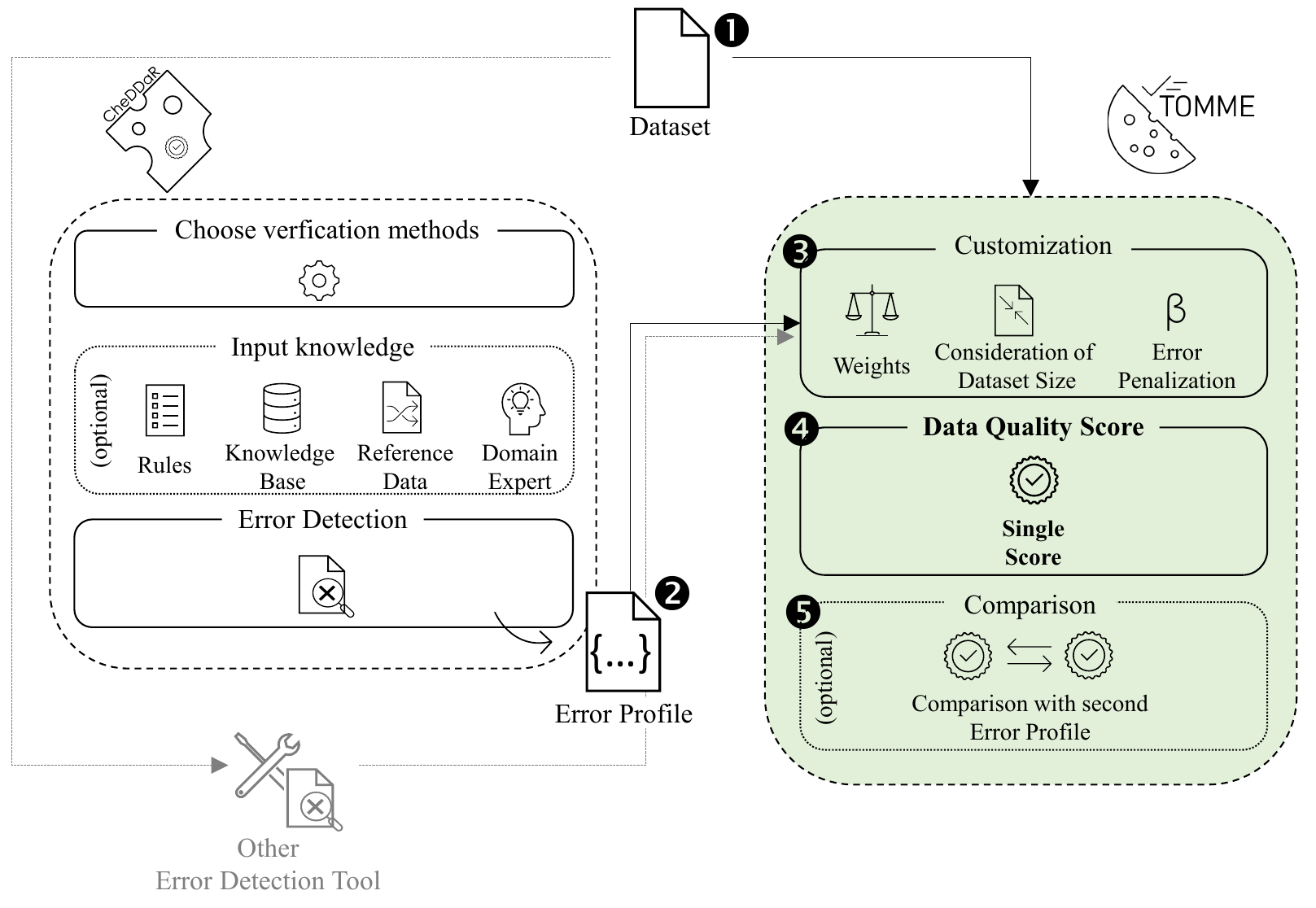}
    \caption{Overview of the architecture and combination of CheDDaR and TOMME}
    %\Description{Picture of architecture overview}
    \label{fig:overview}
\end{figure}

\paragraph{Start} At the beginning~\tikzcircle{1}, the user can input a \ds. %We will provide several \dss for the demonstration, but it is also possible to use own \dss.
The next step is to decide whether to use CheDDaR for error detection or another tool. %However, for the purposes of this demonstration, we will limit it to the use of CheDDaR.
CheDDaR offers several verification options. In the context of TOMME, individual decisions can be made regarding which methods' results should be used. Subsequently, the so-called \textit{error profile} is generated, i.e., the detailed report on the individual error metrics. This error profile is used as input for TOMME~\tikzcircle{2}.

\paragraph{Weights}
As described in the previous section, a central part of TOMME is the individual customization options. These can be configured in the next step~\tikzcircle{3}. The weights for the different properties can be specified in the web UI, as well as the weights for the different error types. In addition, it can be decided whether the number of properties should be taken into account and the penalty factor $\beta$ can be adjusted if necessary -- as described in Section~\ref{sec:ch+t}.

\begin{figure}[ht]
    \centering
    \begin{tikzpicture}
    \node[draw=black, drop shadow, inner sep=0pt] {\includegraphics[width=0.7\linewidth]{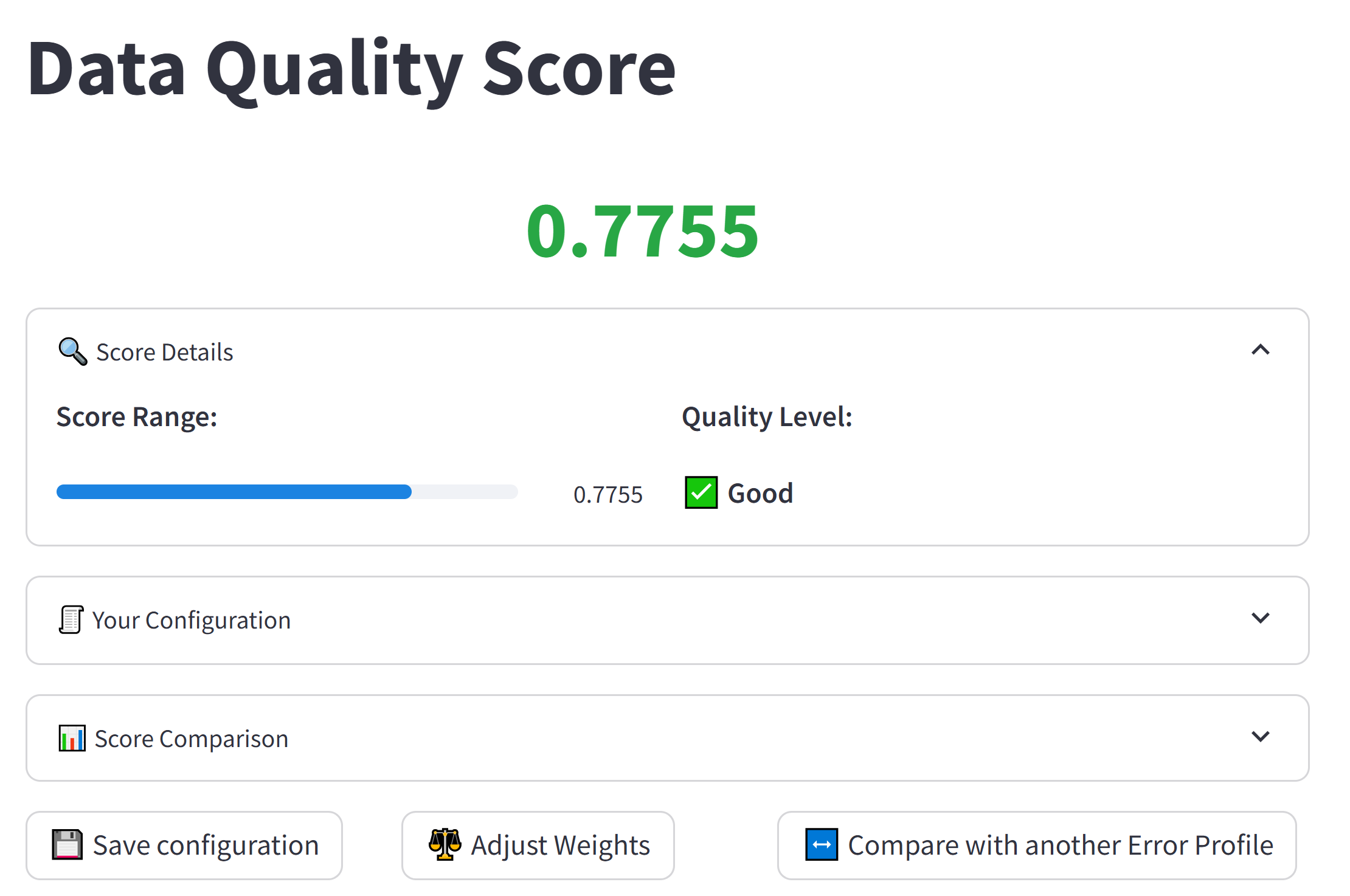}};
    \end{tikzpicture}
    \caption{Data Quality Score -- Screenshot from the web UI}
    %\Description{Screenshot of the data quality score overview}
    \label{fig:dq-score}
\end{figure}

\paragraph{Data Quality Score}
Once all configuration is done, the final data quality score is calculated~\tikzcircle{4}, according to the formula presented in the previous section. The user gets immediate feedback on the quality of the data. In addition, you can view the selected configuration and compare the score with previous calculations. This allows the user to quickly and easily assess the effects of the weights. They can then choose whether they want to save the current configuration, adjust the weights, or compare the score with another error profile in the next step (see Figure~\ref{fig:dq-score}).

\paragraph{Comparison}
If required, another error profile can be uploaded (for example, the result of a slightly modified data cleaning pipeline). The respective data quality scores are compared~\tikzcircle{5} and the user can see which pipeline leads to a better result. It is also possible to adjust the customizations again.

%\paragraph{Interaction and User Engagement}
%In the demonstration, users can explore the data quality of real and synthetic data. They can adjust the numerous configurations and examine the impact on the data quality score. In addition, they can have their own \dss analyzed.

\section{Conclusion and Outlook}
\label{sec:conc}
In this paper, we have introduced TOMME, a generalized, weighted form of accuracy. Using TOMME, it is possible to measure data quality in terms of the \textit{magnitude of error}. In contrast to commonly used accuracy metrics, weights can be applied to properties and error types to refine the measurement for specific application scenarios. Since the formula is not limited to the data item level, it can also be used for semi-structured data, for example. When combined with a detailed error report (such as one that can be generated using CheDDaR), this enables a comprehensive assessment of data quality -- both at a very granular level and using a single score. We have made TOMME publicly available as a web UI\footnote{https://gitlab.com/cheese-board/tomme} so that users can reproduce the experiments and evaluate their own data and data cleaning pipelines.

In future work, we plan to further develop the TOMME metric. First, we want to include further error types and levels. Then, we want to analyze the default weightings for various scenarios to provide users with guidance on how these should be set in practice. Furthermore, we want to analyze what other data quality criteria can be included in the score. In addition to the total number of errors, it might also be interesting, for example, to examine the distributions of the properties before and after data cleaning in order to draw conclusions about the quality of the data cleaning pipeline.

\bibliographystyle{alpha}  
\bibliography{bibliography}  %%% Remove comment to use the external .bib file (using bibtex).

\end{document}